# Deep Learning Aided Rational Design of Oxide Glasses

*R. Ravinder[1], Karthikeya H. Sreedhara[1], Suresh Bishnoi[1], Hargun Singh Grover[1], Mathieu Bauchy[2], Jayadeva[3], Hariprasad Kodamana[4,*], N. M. Anoop Krishnan[1,5,*]*

[1]Department of Civil Engineering, Indian Institute of Technology Delhi, Hauz Khas, New Delhi 110016, India

[2]Physics of AmoRphous and Inorganic Solids Laboratory (PARISlab), Department of Civil and Environmental Engineering, University of California, Los Angeles, CA 90095, USA

[3]Department of Electrical Engineering, Indian Institute of Technology Delhi, Hauz Khas, New Delhi 110016, India

[4]Department of Chemical Engineering, Indian Institute of Technology Delhi, Hauz Khas, New Delhi 110016, India

[5]Department of Materials Science and Engineering, Indian Institute of Technology Delhi, Hauz Khas, New Delhi 110016, India

*Corresponding authors: H. Kodamana (kodamana@iitd.ac.in), N. M. A. Krishnan (krishnan@iitd.ac.in)

**Abstract**

Despite the extensive usage of oxide glasses for a few millennia, the composition–property relationships in these materials still remain poorly understood[1–6]. While empirical and physics-based models have been used to predict properties, these remain limited to a few select compositions or a series of glasses[7–10]. Designing new glasses requires *a priori* knowledge of how the composition of a glass dictates its properties such as stiffness, density, or processability. Thus, accelerated design of glasses for targeted applications remain impeded due to the lack of universal composition–property models[2,3,11]. Herein, using deep learning, we present a methodology for the rational design of oxide glasses. Exploiting a large dataset of glasses comprising of up to 37 oxide components and more than 100,000 glass compositions, we develop high-fidelity deep neural networks[12–14] for the prediction of eight properties that enable the design of glasses[1,2], namely, density, Young's modulus, shear modulus, hardness, glass transition temperature, thermal expansion coefficient, liquidus temperature, and refractive index. These models are by far the most extensive models developed as they cover the entire range of human-made glass compositions. We demonstrate that the models developed here exhibit excellent predictability, ensuring close agreement with experimental observations. Using these models, we develop a series of new design charts, termed as glass selection charts. These charts enable the rational design of functional glasses for targeted applications by identifying unique compositions that satisfy two or more constraints, on both compositions and properties, simultaneously. The generic design approach presented herein could catalyze machine-learning assisted materials design and discovery for a large class of materials including metals, ceramics, and proteins.

**Introduction**
Since their discovery about 5000 years ago, oxide glasses have been an integral part of the human history making them one of the most impactful materials in the world[3,11,15,16]. Apart from their ubiquitous use as automotive wind shields, smart-phone or computer screens, and kitchen-wares, oxide glasses are used for more advanced applications such as nuclear waste immobilization[4,17], energy materials (seals for solid oxide fuel cells[18], electrode or electrolyte materials[19]), and bioactive implants[3,5,11]. Thanks to their compositional flexibility, oxides of nearly any element can form a glass when added to glass-forming oxides such as $SiO_2$, $B_2O_3$, and $P_2O_5$, among others[1]. This makes the possible compositions of oxide glasses to be more than $\sim 10^{26}$, considering a 1 mol% increment for each of the oxide components in the glass[1]. Despite such wide usage of glasses, the understanding of composition–property relationships in glasses have been limited to a small subset of glass compositions[1,7,8], leading to an Edisonian trial-and-error methodology for glass discovery[2]. This impedes the accelerated development of new glass compositions for targeted applications due to the following reasons. (i) Traditional trial-and-error method is extremely cost, time, and resource intensive. (ii) Due to the large number of glass compositions possible, along with their extremely nonlinear composition–property relationships, the number of experiments or simulations required to explore the entire compositional space is astronomical[1–3]. (iii) Design of a new glass is never based on a single target property—at the very minimum, a glass also needs to be processable, posing constraints on its liquidus temperature and glass transition temperature. (iv) Obtaining multiple properties that are relevant for glass design for the same composition requires additional simulations or experiments, and hence, are sparsely reported in the literature. (v) Discovering an optimum composition has thus far relied mostly on experience-based intuition and serendipity—it is not guaranteed that the obtained result represents a true optimal composition. This calls for the development of an alternate scalable and robust approach for accelerating the design and discovery of glasses[2].

Extensive experimental studies in the past century have been documented in various glass databases[6,20,21]. Recent studies on data-driven modelling, exploiting these databases, have shown to be indeed promising to predict the composition–property relationships in glasses[2,22–25] and other materials[26–29]. However, most of the studies on glasses are limited to a small series of glasses with focus on one or two properties[22,24,25,30] making them restrictive and, in practice, unusable for glass design, where a combination of properties are desired. To address this challenging problem, we propose a rational design paradigm (see Figure 1) based on deep learning (DL) to develop universal models that provides two-way maps of the composition–property space of oxide glasses. Specifically, we choose 37 oxide components and more than 100,000 glass compositions encompassing glasses used for a wide range of commercial applications. These datasets are used to develop universal composition–property DL models for eight key properties, namely, density, Young's modulus, shear modulus, Vickers hardness, glass transition temperature ($T_g$), thermal expansion coefficient (TEC), liquidus temperature, and refractive index. These properties are highly relevant for glass design—for instance, maximizing the Young's modulus to density ratio along with maximum hardness is a criterion that is relevant for developing light-weight, scratch-resistant glasses[1,31,32]. The design can be further enhanced by imposing constraints on thermal expansion coefficient and liquidus temperature, so as to avoid thermal shock while controlling the processability, respectively. It should be noted that the properties of glass depend on its thermal and pressure history[33], owing to the non-equilibrium nature of glasses[1,34]. Further, some properties such as hardness and glass transition temperature may also depend on the definition and the technique used to measure[1,32,33]. To maintain consistency, we use the dataset on glasses that are cooled with the experimental cooling rates. Further, we rely on standard definitions of the properties[1] and specifically, for hardness, we use the Vickers hardness values[1].

The DL models reported here are by far the most extensive models, covering the entire range space of the selected compositions. These models are then used to: (i) populate the composition–property space for multiple properties in multicomponent glasses that are otherwise unavailable from the literature, (ii) understand the composition-dependent variations of glass properties while capturing the underlying physics (for instance, modifier effects as we show later), and (iii) develop multi-property design charts termed as glass selection charts (GSCs) enabling rational design of glasses. Finally, to accelerate the design of glasses, we developed a software package incorporating the rational design paradigm, namely, Python for Glass Genomics (PyGGi, see http://pyggi.iitd.ac.in), which is made available publicly.

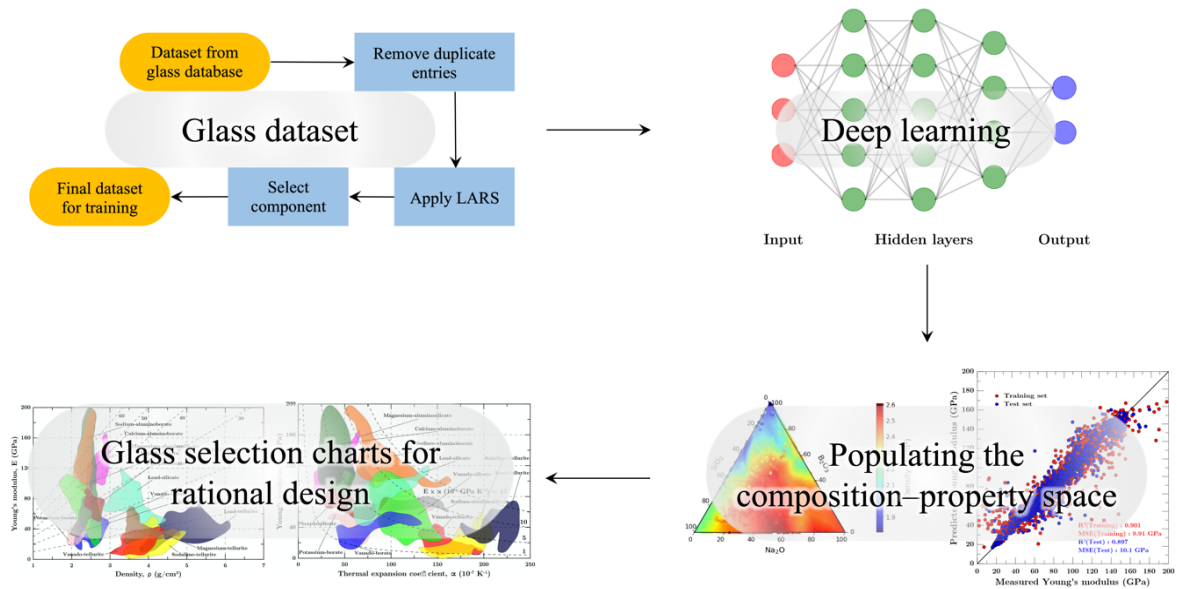

**Figure 1. Schematic representation of the methodology for the rational design of glasses.** Glass dataset comprising of the input compositions and the corresponding properties are collected from the literature and glass databases. Deep learning (DL) models developed based on the processed dataset are used to populate the composition–property space and develop glass selection charts. These charts act as the road-map to selecting compositions satisfying multiple constraints, thus, enabling the design of novel glasses for targeted applications.

**Methods**
   (i)    **Data preparation**
The initial dataset of glasses was taken from the literature and glass datasets such as INTERGLAD V7.0 and SciGlass, which is a database of the experimental data for more than 350,000 glasses. Note that the glass databases are a compilation of the experimental data from the literature with the complete annotation including the journal, authors etc. The initial raw data consisted of duplicate and incomplete entries, which were removed. Further, some inconsistent compositions (for example, the compositional sum not adding up to 100%) were also removed. The clean data consisted of glasses with more than 300 features (oxide components), with most of the features having very few entries. The features with very less number of data entries lead to poor model training. Therefore, the input features were further reduced using dimensionality reduction algorithm, least angle regression (LARS). We selected important features for each property and took an intersection of all the important features, which resulted in 37 features. Final dataset for training was prepared using these 37 features, namely, $SiO_2$, $B_2O_3$, $Al_2O_3$, MgO, CaO, BaO, $Li_2O$, $Na_2O$, $K_2O$, $Ag_2O$, $Cs_2O$, $Tl_2O$, BeO, NiO, CuO, ZnO, CdO, PbO, $Ga_2O_3$, $Y_2O_3$, $La_2O_3$, $Gd_2O_3$, $Bi_2O_3$, $TiO_2$, $ZrO_2$, $TeO_2$, $P_2O_5$, $V_2O_5$, $Nb_2O_5$, $Ta_2O_5$, $MoO_3$, $WO_3$, $H_2O$, $Sm_2O_3$, $MgF_2$, $PbF_2$, $PbCl_2$. This dataset includes multicomponent glasses with the components varying from two to fifteen.

### (ii) Model Development

To develop the deep feedforward neural nets[12,13], we relied on the scikit-learn python library. The cleaned dataset was used with the molar composition as the input and the relevant property as the output. A split of 70:30 was used for splitting the training and test data. Deep learning (DL) models were developed for each of the properties separately. To this extent, the performance of the training set was maximized against various performance objectives such as root mean-squared error (RMSE), $R^2$, and mean absolute error. Final model set selections were done based on the RMSE. While training, various activations functions were employed in the DL model and finally, rectified linear unit (ReLU) was chosen as the candidate activation function due to its superior performance. To optimize the model structure, we employed a grid search approach, wherein the both the number of hidden layers and units per hidden layer were varied systematically. The structure that exhibited maximum performance was chosen as the candidate for further refinement using hyper-parametric optimization. A five-fold cross validation was employed to optimize the final structure of the DL model. Hyper-parameters[12,14] such as learning rate, error norms, number of units per hidden layer, and solvers (l-BFGS, SGD, ADAM) were further optimized to obtain the final refined model. These final models were used to obtain the composition–property maps and the glass selection charts.

**Results and Discussions**

In order to develop composition–property DL models, we employ the following steps, the schematic of which is presented in Figure 1. (i) First, we clean the dataset and perform least angle regression (LARS) to choose the input components (from ~300 to 37 oxide components in this case), which form the most commonly available glass compositions. LARS ensure that the components chosen here are well-represented in the final dataset. Note that the number of glass compositions available for training the DL models vary depending on the specific properties of interest. (ii) Second, we use the "clean" dataset for the supervised DL model, specifically, deep feedforward neural nets[12] for each properties, separately. The input features for the DL models are only the glass compositions, while the outputs, against which the model is trained, are the glasses' properties. (iii) Third, the final structure of the DL models are further optimized using hyperparametric optimization[13,14] to reduce the model complexity (see Methods). (iv) Finally, the predictive models are used to develop GSCs.

Figure 2 shows the predictions of the optimized DL models for the training (70%) and test (30%) dataset corresponding to the eight properties of oxide glasses. Specifically, Figures 2(a)–(h) shows the predicted values of density, Young's modulus, hardness, shear modulus, TEC, $T_g$, liquidus temperature, and refractive index, respectively, with respect to the experimental values. Due to the large number of overlapping values, a heatmap with the color representing the number of points per unit area is used. The inset show the error in the predicted values as a probability density function with the shaded region representing the 90% confidence interval. Overall, the DL models presented here exhibit highly accurate predictions of the properties with respect to experimental values for a wide range of oxide glasses with up to 37 input components (see Fig. 2(a)–(h)).

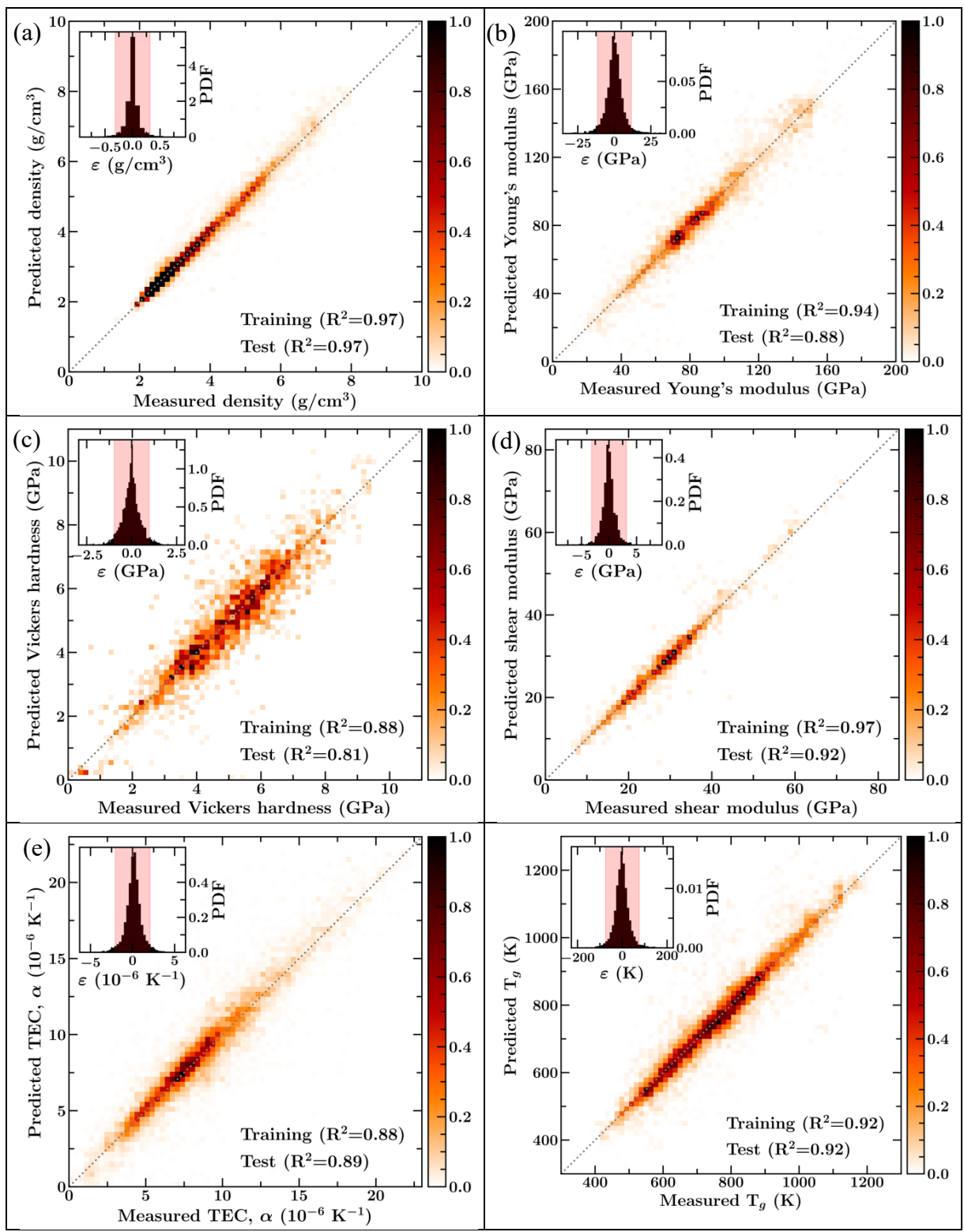

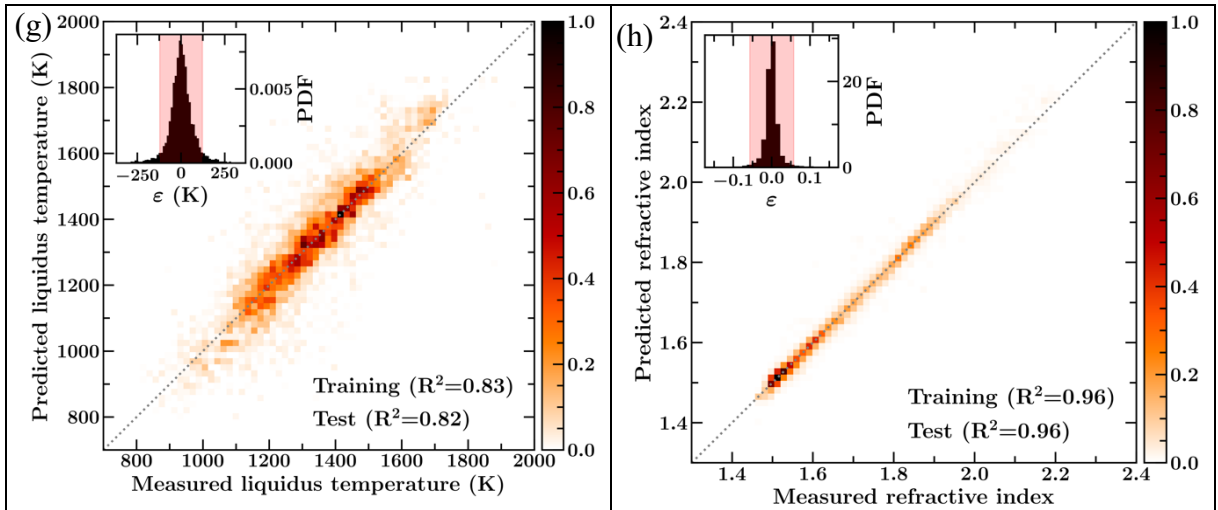

**Figure 2. Property predictions based on deep learning (DL) models.** Predicted values of **(a)** density, **(b)** Young's modulus, **(c)** hardness, **(d)** shear modulus, **(e)** thermal expansion coefficient (TEC), **(f)** glass transition temperature ($T_g$), **(g)** liquidus temperature, and **(h)** refractive index of oxide glasses using optimized DL models with respect to the experimental values. Note that the color represents the number of points per unit area for associated with each property following the respective coloring scheme. The inset shows the error in the predicted values as a probability density function (PDF) with the shaded region representing the 90% confidence interval. $R^2$ values of the training and test set are also provided.

In order to demonstrate the capability of the models to explore the complete compositional space, we predict the properties of a ternary calcium aluminosilicate glasses, that is, $(CaO)_x(Al_2O_3)_y(SiO_2)_{1-x-y} \forall (x+y) \leq 1; x, y \in [0,1]$. Figures 3(a)–(h) shows the ternary plots for density, Young's modulus, hardness, shear modulus, TEC, $T_g$, liquidus temperature, and refractive index, respectively, of calcium aluminosilicate glasses. The experimental compositions are presented by square markers with the heatmap representing the values of the respective property. In addition, the underlying heatmap in the ternary diagram shows the predicted values using the DL models for each of the respective properties. As we observe from Figure 3(a)–(h), the predicted values for all properties exhibit a close agreement with the experimental values for the selected ternary. Further, the model provides access to the complete compositional domain (in this case the entire ternary space), thereby predicting the properties of hitherto unknown compositions.

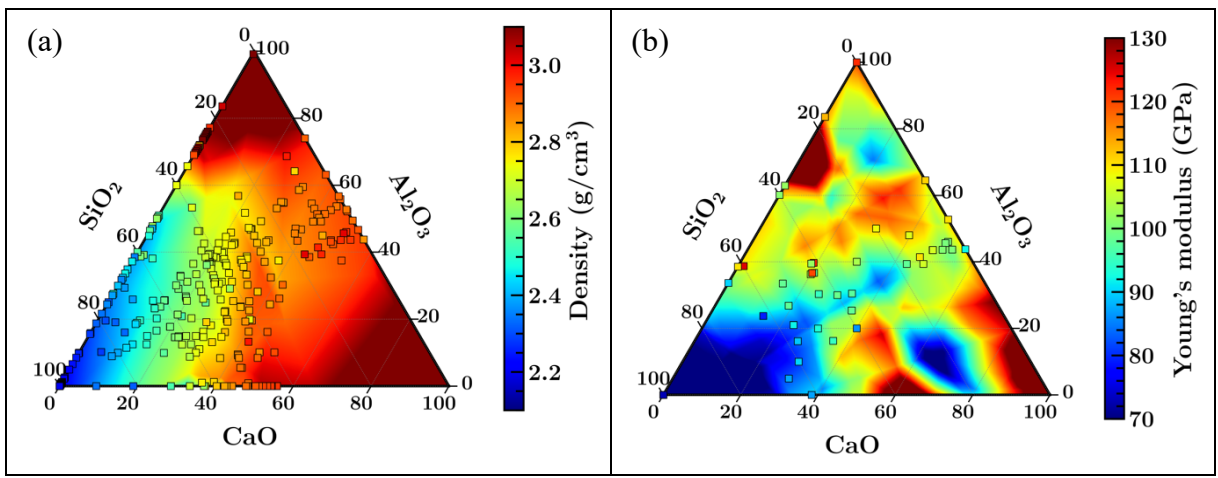

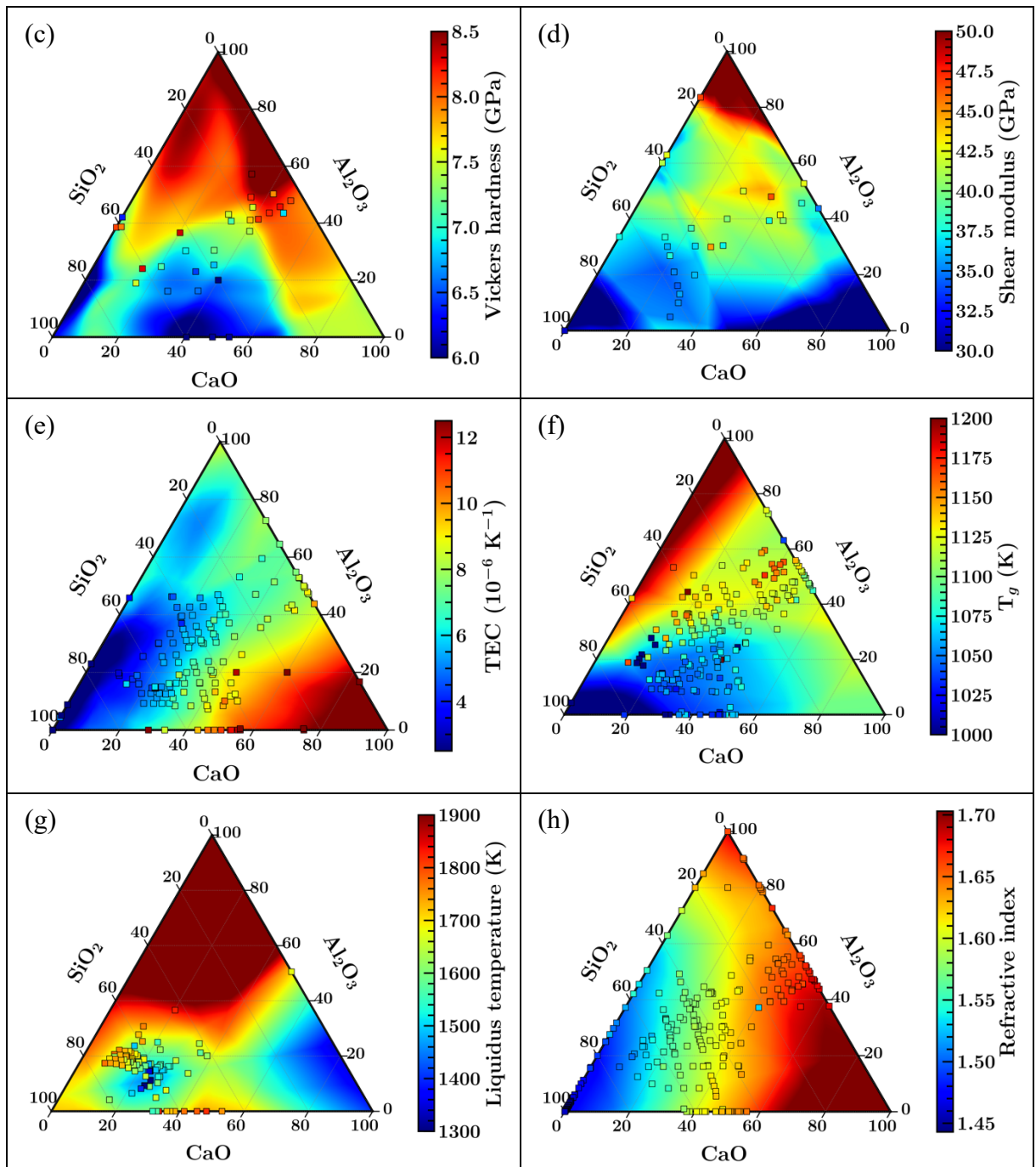

**Figure 3. Ternary plots of calcium aluminosilicate glasses from DL models.** Ternary plots of **(a)** density, **(b)** Young's modulus, **(c)** hardness, **(d)** shear modulus, **(e)** thermal expansion coefficient (TEC), **(f)** glass transition temperature ($T_g$), **(g)** liquidus temperature, and **(h)** refractive index of calcium aluminosilicate glasses. Square markers represent experimental compositions with the color representing the value of the property from the experiments. The underlying heatmap represents the predictions of the model. The scale bar corresponding to each subfigure shows the property range along with the coloring scheme.

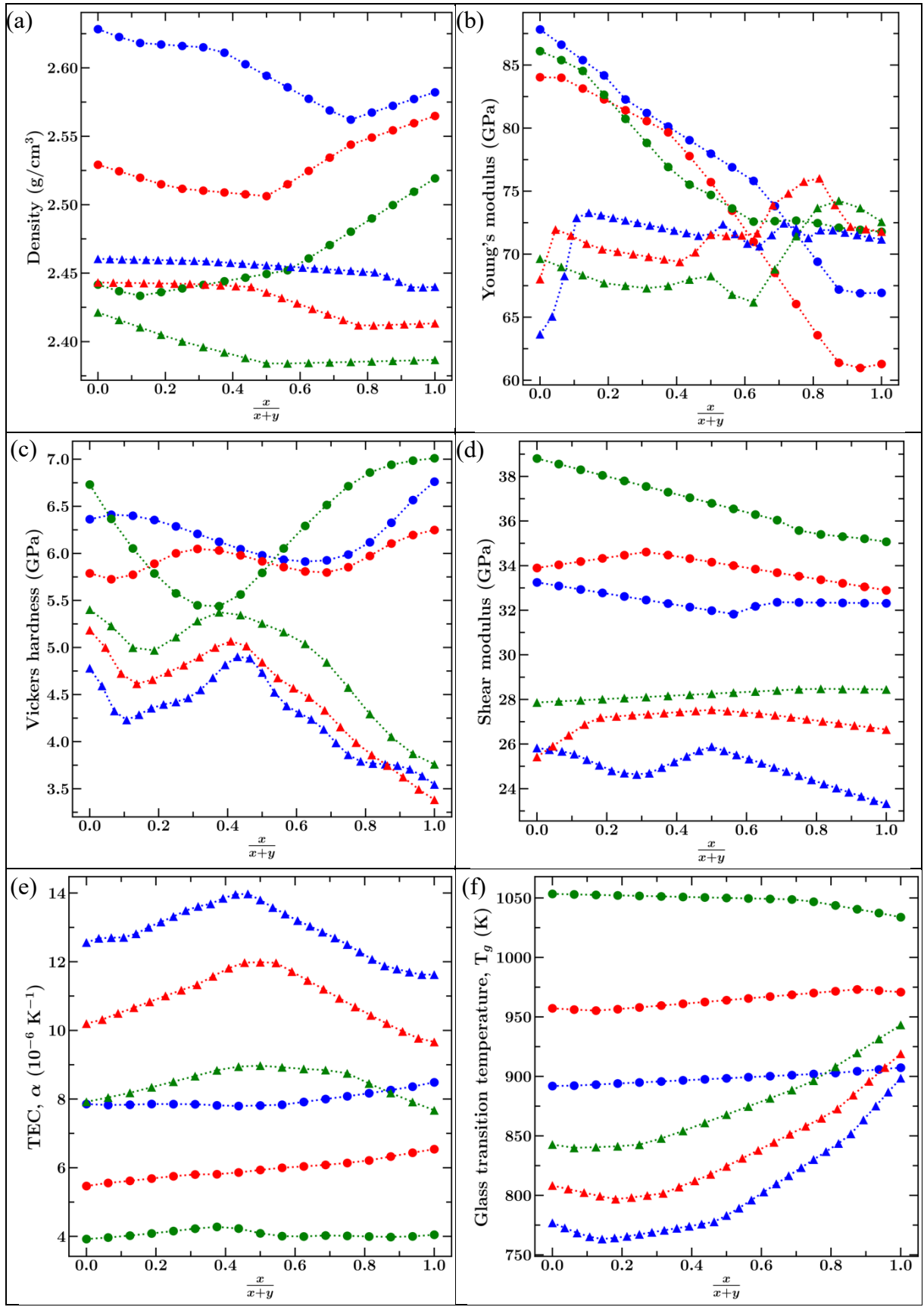

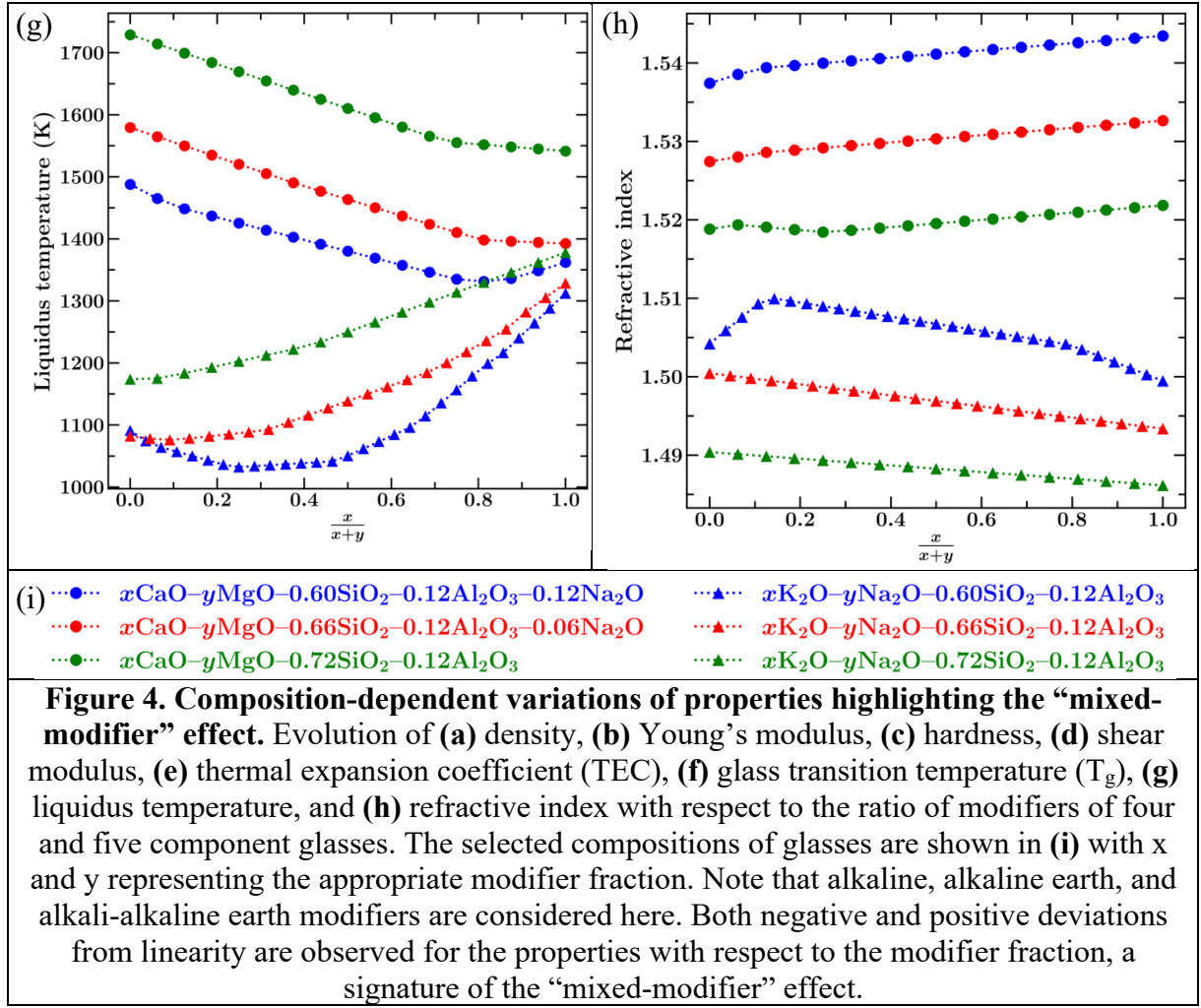

**Figure 4. Composition-dependent variations of properties highlighting the "mixed-modifier" effect.** Evolution of **(a)** density, **(b)** Young's modulus, **(c)** hardness, **(d)** shear modulus, **(e)** thermal expansion coefficient (TEC), **(f)** glass transition temperature ($T_g$), **(g)** liquidus temperature, and **(h)** refractive index with respect to the ratio of modifiers of four and five component glasses. The selected compositions of glasses are shown in **(i)** with x and y representing the appropriate modifier fraction. Note that alkaline, alkaline earth, and alkali-alkaline earth modifiers are considered here. Both negative and positive deviations from linearity are observed for the properties with respect to the modifier fraction, a signature of the "mixed-modifier" effect.

Now, we proceed to demonstrate the capability of the DL models to predict the extremely nonlinear composition-dependent variations of properties. It is well-known that glasses with mixed modifiers (for e.g., alkali or alkaline earth cations) exhibit non-additive, nonlinear trends in their properties such as density, Young's modulus, glass transition temperature, and hardness—a manifestation commonly known as the "mixed-modifier effect"[32,35–37]. To analyse whether the models developed herein capture this mixed-modifier effect, we consider a series of aluminosilicate glasses with alkali and alkaline earth modifiers. Specifically, the glass compositions considered consists of the aluminosilicate network and different combinations of $Na_2O$, $K_2O$, $CaO$, and $MgO$ as the network modifiers (see Fig. 4). Figure 4(a)–(h) shows the variations of density, Young's modulus, hardness, shear modulus, TEC, $T_g$, liquidus temperature, and refractive index, respectively, respectively, of the selected glasses. Six series of glasses are considered with different modifiers such as $Na_2O$, $K_2O$, $CaO$, $MgO$ added in different ratios (see Fig. 4(i)). Interestingly, we observe that the DL models show significant negative and positive deviations from the linear additive behaviour for all the properties in agreement with the experimental observations[32,35,36]. Specifically, the highly nonlinear trend of hardness with respect to increasing CaO content is noteworthy—a topic of extensive research in previous experimental and theoretical studies on the mixed modifer effect[32]. It is worth noting that the DL models span the compositional range with higher resolution (by interpolating the data points within the dataset), thereby allowing a finer inspection of the underlying physical trend. Although the trends presented herein may also occur due to the model variance, the rigorous hyperparametric optimisation carried out during the training phase minimizes its effect significantly. Further, an optimized DL model captures the intrinsic relationship between composition and properties, probably more than the experimental data themselves. Thus, the DL models also allow one to understand the composition–property trends, thereby providing insights into the fundamental material characteristics.

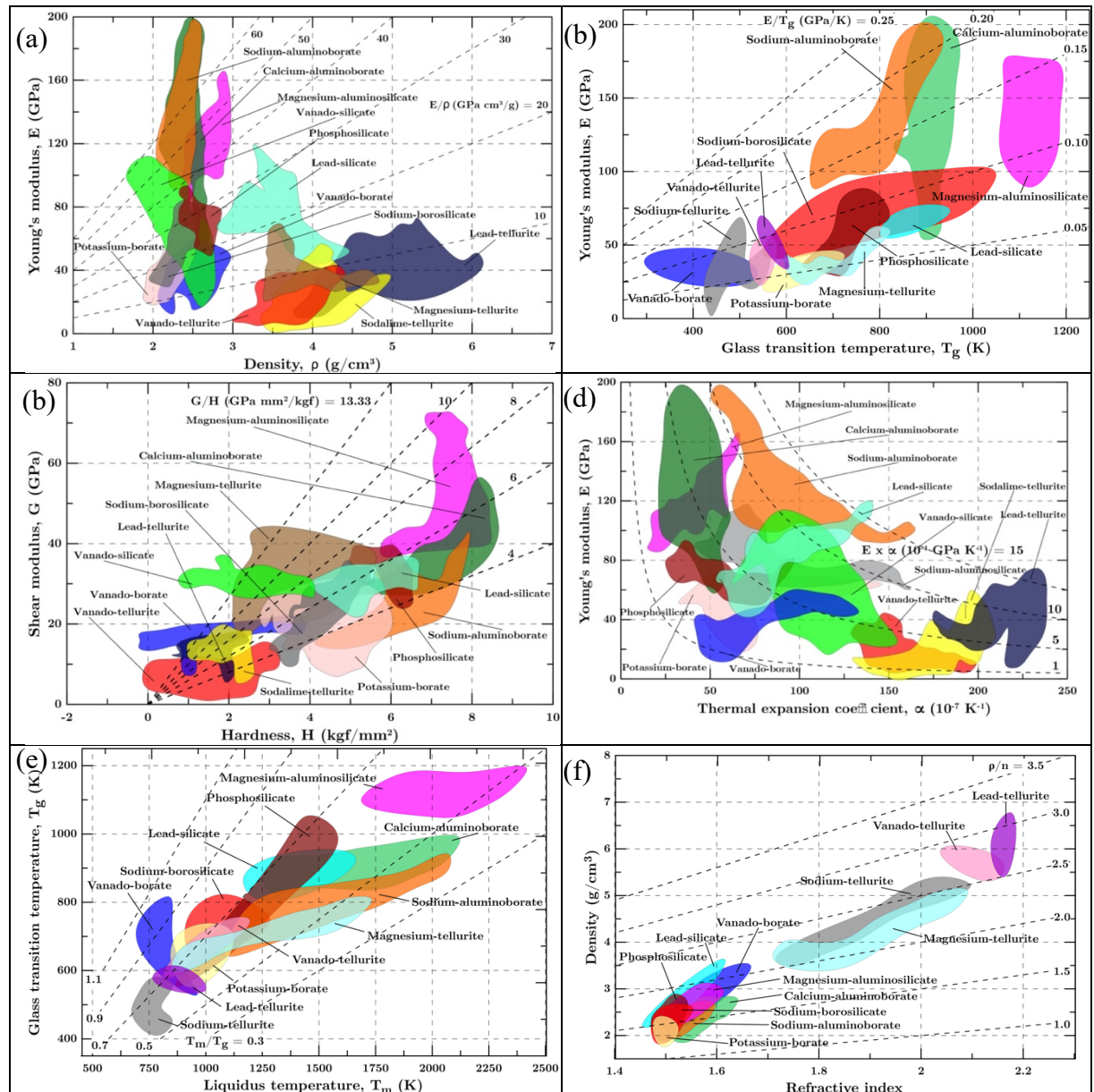

**Figure 5. Glass selection charts.** Variation of two independent properties for a large series of commonly used glasses, namely, aluminosilicate, silicate, borate, borosilicate, phosphosilicate, aluminoborate, and tellurite. **(a)** Young's modulus with density, **(b)** Young's modulus with glass transition temperature, **(c)** shear modulus with hardness, **(d)** Young's modulus with thermal expansion coefficient, **(e)** Glass transition temperature with liquidus temperature, and **(f)** density with refractive index. Dotted lines in the figures correspond to design lines with X/Y = constant for subfigures (a), (b), (c), (e), and (f) and XY = constant for subfigure (d). These glass selection charts can enable the rational design of glasses by allowing the choice of glasses subjected to multiple property and compositional constraints.

Finally, we go on to develop a series of multi-property design charts with a view to aid the rational design of glasses for targeted applications. These design charts, termed as GSCs, represent a large body of multivariate multi-property information which, when interpreted appropriately, allows accelerated and economical design of glasses. Here, we draw parallels for the GSCs to the material selection charts, also known as Ashby plots[38,39]. While these GSCs could be high dimensional, for illustrative purposes, we limit ourselves to two-dimensional GSCs. In these charts, we plot the variation of two independent properties for a large number of multicomponent glasses with different network formers and modifiers (approximately 60,000 compositions in this case). Additional compositions or properties can be included as per the design interests and requirements.

Figures 5(a)–(f) show the GSCs of Young's modulus with density, Young's modulus with glass transition temperature, shear modulus with hardness, Young's modulus with thermal expansion coefficient, glass transition temperature with liquidus temperature, and density with refractive index, respectively. These GSCs allow the selection of glass compositions exhibiting a preferred combination of two or more properties in conjunction with the domain expert knowledge[40]. For instance, Fig. 5(a) allows to identify glass compositions exhibiting maximum stiffness to density ($E/\rho$) ratio, a preferred feature for the design of lightweight, yet stiff structures[41]. Thus, the contour plot of $E/\rho$ provides the glass compositions having an identical value of specific modulus. It should be noted from Fig. 5(a), that for glasses with a constant specific modulus of 10 GPa.cm$^3$/g, the density ranges from 2 to 5.5 g/cm$^3$ allowing a rational choice for glasses in applications such as optical fiber, or structural glasses[42]. Similarly, designing glass composites with minimal or zero thermal stress requires the choice of compositions having similar $E\alpha$ values, to avoid any eigen stress due to the mismatch in the thermal stresses. $E\alpha$ represents the thermal stress developed in a material associated with a unit change in temperature. Thus, the contour line with constant $E\alpha$ value (see Fig. 5(d)) provides the list of all the glass compositions that develop identical thermal stress associated with a temperature change. A contour line with an $E\alpha$ value of 5 GPa/K yields a spectrum of glass compositions with the Young's modulus ranging from 20 to 180 GPa and thermal expansion coefficient ranging from 25 to 240 $\times$ 10$^{-7}$/K. Similar design choices could be drawn from other GSCs as well for a variety of applications such as scratch resistant glasses, glasses for nuclear waste immbolization, and optical fibres. Further, the GSCs also provide insights into the correlations of different properties, for example, a positive correlation with Young's modulus and glass transition temperature[43], or shear modulus and hardness[44], the atomic origin of which requires to be investigated using tailored experiments and simulations. Similarly, the ratio of $T_g$ and liquidus temperature may provide insights to glass forming ability[45]. Thus, Fig. 5(f) can provide a guide to identifying compositions that can easily form a glass.

**Conclusion**

Overall, the present work demonstrates the importance of employing DL for accelerated design and discovery of oxide glasses. The scalable approach presented herein may be applied to a large class of materials ranging from crystals to disordered materials. In such cases, additional features governing the properties such as processing parameters, microstructures, and phases can enhance the reliability and accuracy of these models, if incorporated appropriately. The models can also be used for designing optimal routes for material selection and synthesis through surrogate model based optimizations. Altogether, the rational design approach combining data-driven modelling and expert knowledge can open new vistas for the development next generation tailor-made materials.